\documentclass[aoas,nameyear,dvips]{arximspdf}

\doi{10.1214/10-AOAS381}
\volume{4}
\issue{4}
\pubyear{2010}
\firstpage{1638}
\lastpage{1641}

\makeatletter
\newcommand{\EE}{\mathbb{E}}
\makeatother

\begin{document}
\begin{frontmatter}

\title{Remembrance of Leo Breiman}
\runtitle{Remembrance of Leo Breiman}

\begin{aug}
\author[A]{\fnms{Peter} \snm{B\"uhlmann}\ead[label=e1]{buhlmann@stat.math.ethz.ch}\corref{}}
\runauthor{P. B\"uhlmann}
\affiliation{ETH Z\"urich}
\address[A]{Seminar f\"ur Statistik\\
ETH Zentrum, HG G17\\
CH-8092 Zurich\\
Switzerland\\
\printead{e1}} %adresu isvedimo komanda gale!
\end{aug}

% HISTORY:
\received{\smonth{7} \syear{2010}}

% ABSTRACT

% KEYWORDS
%


\end{frontmatter}

%s1 ###
\section{How I met Leo Breiman}

In 1994, I came to Berkeley and was fortunate to stay there three years,
first as a postdoctoral researcher and then as Neyman Visiting Assistant
Professor. For me, this period was a unique opportunity to see other
aspects and learn many more things about statistics: the Department of
Statistics at Berkeley was much bigger and hence broader than my home at
ETH Z\"urich and I enjoyed very much that the science was perhaps a bit more
speculative.

As soon as I settled in the department, I
tried to get in touch with the local faculty. Leo Breiman started a reading
group on topics in machine learning and I didn't hesitate to participate
together with other Ph.D. students. Leo spread a tremendous amount of
enthusiasm, telling us about the vast opportunity we now had by taking
advantage of computational power. Hearing his views and opinions and
listening to his thoughts and ideas has been very exciting, stimulating
and entertaining as well. This was my first occasion to get to know
Leo. And there was, at
least a bit, a vice-versa implication: now, Leo knew my name
and who I am. Whenever we saw each
other on the 4th floor in Evans Hall, I got a very gentle smile and
``hello'' from Leo. And in fact, this happened quite often: I often walked
around while thinking about a problem, and it seemed to me, that
Leo had a similar habit.

%s2 ###
\section{Witnessing three of Leo's fundamental contributions}

I only got to know Leo Breiman in his late career. Nevertheless, between
1994 and 1997 when I was in Berkeley, I could witness Leo's exceptional
creativity when he invented Bagging [\citet{brei96}], gave fundamental
explanations about Boosting [\citet{brei99}] and
started to develop Random Forests [\citet{brei01}].

%s2.1 ###
\subsection{Bagging}
I had the unique opportunity to listen to Leo Breiman when he presented
Bagging during a seminar talk at UC Berkeley. I was puzzled and
intrigued. At that time, I was working on the bootstrap and what Leo
said didn't sound right to me: using the bootstrap language, he
proposed to use
$\hat{\theta}_{\mathrm{Bag}} = \EE^*[\hat{\theta}^*]$,
where $\hat{\theta}$ is the output of a ``complex algorithm'' based
on the
original observations and $\hat{\theta}^*$ denoting the analogue
based on
the bootstrap sample. Trivially,
\[
\hat{\theta}_{\mathrm{Bag}} = \hat{\theta} + (\EE^*[\hat{\theta
}^*] - \hat{\theta}),
\]
and hence from this point of view, Leo has proposed to use the original
estimator and \textit{adding} the classical bootstrap bias correction
estimate (instead of subtracting it). But this is not an appropriate view
for the problem Leo was looking at, and---as usual---it turned out that he
was right. Of course, Leo didn't present Bagging in this way: he argued via
stability [\citeauthor{brei96} (\citeyear{brei96}, \citeyear{brei96b})] and
that unstable estimators can be stabilized using the
bootstrap. I still remember how Leo presented during the seminar talk
many empirical examples, one batch of datasets after the other,
demonstrating that Bagging
improves the prediction performance by about 30\%. It was great news! And
also a kind of shock that nobody among the people in the audience or in the
community had thought about it before.

After the seminar, I tried it out myself: it's so simple and easy to do!
And indeed, Bagging worked when using CART trees or other ``unstable''
procedures. And in terms of prediction, it didn't do any harm for
``stable'' procedures. I have been fascinated by the idea, I started
working on it and eventually, Bin Yu and I had some additional
explanations why
Bagging works [\citet{pbyu02}]---a~tiny contribution in comparison
to Leo's breakthrough.

%s2.2 ###
\subsection{Arcing and Boosting}

In 1996, \citet{freusch96} published their AdaBoost
algorithm and
they showed many empirical examples where their method performed exceptionally
well. This caught a lot of attention, and maybe even more so than with
Bagging, one wondered why such an ensemble method based on mysterious
re-weighting works so well. Leo Breiman also got involved: he proposed a
variant of Boosting called ``Arcing'' [\citet{brei98}] and then
once more, he
made a breakthrough: he formalized AdaBoost as a gradient descent
optimization in function space where the gradient is estimated by a
nonparametric procedure such as a CART tree [\citet{brei99}]. Many people,
particularly from statistics, followed up on Leo's formalized framework
[\citet{fht00}; \citet{pbyu00}; \citet{fried01}; \citet{pbyu03}].
Part of my own research has built up on this result and Leo's result
had a big and crucial influence on my research.

Leo's important
and deep contribution in Boosting was about understanding the
algorithm and not in terms of developing a new method. Maybe this was an
interesting ``outlier'' in Leo's late career where he primarily was the
designer of new methods and algorithms. But it fits perfectly into the
picture: my remembrance of Leo is not only about his outstanding
creativity but also about his analytical thinking regarding algorithms and
machine learning---which is not a complete surprise given his mathematical
background and training.

%s2.3 ###
\subsection{Random Forests}

A third fundamental contribution of Leo's late career is the
development of
Random Forests, and I have a special memory on this. I was at home in
Switzerland and Don Geman gave a talk at ETH
Z\"urich about using trees with randomization at the nodes [\citet{amitg97}]. I
spoke with Don and told him about Leo's Bagging which randomizes the
samples instead of the nodes in the tree but Don was convinced that
node randomization is much better. Leo took this suggestion and carried it
much further. In particular, he created the idea of incorporating ``variable
importance,'' knowing well in advance
that people will use it in complex data problems with thousands of
variables as in, for example, high-throughput molecular biology
[\citet{uri06}; \citet{fred09}].

Random Forests is an astonishingly powerful ``off-the-shelf''
method.\break Whether we like such ``off-the-shelf'' procedures or not,
Random Forests
works extremely well in great generality, given that it is a pure machine
learning algorithm which
essentially does not even require the specification of a tuning parameter!
There is virtually no competing method which can so easily deal with
high-dimensional continuous,
categorical or mixed data yielding powerful
predictions and some ``first-order'' information about
variable importance. There have been some attempts for better
(mathematical) understanding of Leo's Random Forests [\citet{lin06}; \citet{biau08}] and
I tried myself some years ago. However, without having Leo's deep
insights and
intuition, it's maybe still a bit of a mystery why Random Forests works so
well.

%s3 ###
\section{Being influenced by Leo}

Leo's grand views, visions and his research had a profound influence on my
own scientific life. My joint work with Adi Wyner on
``Variable Length Markov Chains'' [\citet{pbwy99}], developed
during my time
in Berkeley, is a tree model and certainly inspired by CART
[\citet{brei84}]. Similarly, a tree-based GARCH model with
Francesco Audrino
[\citet{audpb01}] is an adaptation of CART. Much more
obvious is the connection of my joint work with Bin Yu on
Bagging and Boosting [\citeauthor{pbyu00} (\citeyear{pbyu00}, \citeyear{pbyu02}, \citeyear{pbyu03}, \citeyear{pbyu05})]: it was Leo's
excitement and his great ideas that stimulated my curiosity and my interest
in these techniques and more generally in machine learning. My latest
example is
some joint work with Nicolai Meinshausen: what we call ``Stability Selection''
[\citet{mebu10}] is Leo Breiman's idea of Bagging, transferred from
the problem of
making predictions to the notion of variable and feature selection.

Leo Breiman, the pioneer of statistical machine learning: without him, my
scientific life would have gone a different way, and I am tremendously
thankful that I had the chance to know him personally.

\section*{Acknowledgments}
I would like to thank Fred Hamprecht and Markus Kalisch for thoughtful
comments.

%suskaldyti doi

\printaddresses

\end{document}